\documentclass[a4paper]{spie}
\usepackage[dvips]{graphicx}

\title{IIR Adaptive Filters for Detection of Gravitational Waves from Coalescing Binaries}
\author{F. Acernese\supit{a}\supit{b}, F. Barone\supit{c},R. De Rosa\supit{a}\supit{b}, 
A Eleuteri\supit{a}\supit{b}, L Giordano\supit{b}\supit{d} and L Milano\supit{a}\supit{b}
\skiplinehalf
\supit{a} Dip. di Scienze Fisiche, Universit\`a di Napoli ''Federico II'', via Cintia, I-80126 Napoli, Italia \\
\supit{b} INFN, sez. Napoli, via Cintia, I-80126 Napoli, Italia \\
\supit{c} Dip. di Scienze Farmaceutiche, Universit\`a di Salerno, via Ponte Don Melillo, 84084 Fisciano (SA), Italia \\
\supit{d} Dip. di Matematica ed Applicazioni ''R. Caccioppoli'', Universit\`a di Napoli ''Federico II'', via Cintia I-80126 Napoli, Italia
}

\authorinfo{Further author information: send correspondence to Fausto Acernese, e-mail: fausto.acernese@na.infn.it}

\begin{document}
\maketitle

\begin{abstract}
In this paper we propose a new strategy for gravitational waves detection from coalescing binaries, using IIR Adaptive Line Enhancer (ALE) filters. This strategy is a classical hierarchical strategy in which the ALE filters have the role of triggers, used to select data chunks which may contain gravitational events, to be further analyzed with more refined optimal techniques, like the the classical Matched Filter Technique. After a direct comparison of the performances of ALE filters with the Wiener-Komolgoroff optimum filters (matched filters), necessary to discuss their performance and to evaluate the statistical limitation in their use as triggers, we performed a series of tests, demonstrating that these filters are quite promising both for the relatively small computational power needed and for the robustness of the algorithms used. The performed tests have shown a weak point of ALE filters, that we fixed by introducing a further strategy, based on a dynamic bank of ALE filters, running simultaneously, but started after fixed delay times. The results of this global trigger strategy seems to be very promising, and can be already used in the present interferometers, since it has the great advantage of requiring a quite small computational power and can easily run in real-time, in parallel with other data analysis algorithms.
\end{abstract}

\keywords{Gravitational Waves, Interferometers, Data Analysis, Adaptive Filters}

\section{Introduction}
Gravitational Wave (hereafter GW) detection is certainly one of the most challenging goals for today physics: a very strong proof in favor of the Einstein General Relativity description of phenomena related to the dynamics of gravitation and the opening of a completely new channel of information on astrophysical objects\cite{MISNER,BLAIR,SAULSON}. Ground-based and space detectors are now operational or will be operational in the next years, with different bands and sensitivities. The VIRGO/LIGO\cite{VIRGO,LIGO} network of ground-based kilometer-scale laser interferometer gravitational wave detectors will probably be the key to open up a new astronomical channel of information in the frequency band $10\,Hz$ to $10\,kHz$. In addition, when the proposed 5 million kilometers long space based interferometer LISA\cite{LISA} flies, another window will be opened in the frequency band $10^{-5}\div 1\,Hz$. In particular, the sensitivities of the operational interferometers are becoming very interesting, producing reasonable hopes in the researchers that the first detection of gravitational waves is becoming very close. Therefore, the efforts in studying and developing data analysis methodologies and techniques are increasing, such as the tests performed on real data (interferometers outputs), allowing the researchers testing not only the optimality but also robustness and feasibility of such techniques. Within this research context, we focused our attention on GW signal detection from coalescing compact binaries (neutron stars (NS-NS), black holes (BH-BH) or a mixed (NS-BH)), very promising transient GW sources, which may also allow precise measurements of the masses of the objects, of the spins and, in the case of neutron stars, of the radii, too\cite{SCHUTZ}. But although the interferometers seem to be sensitive enough to detect this class of sources, nevertheless the problem of GW signal analysis has not yet globally solved. For this task many solutions are being evaluated, in particular for what concerns the choice of suitable data analysis techniques taking into full account the shape of the expected signal, the noise of the detector and the available computing power\cite{THORNE}.

Many efforts have been made for the development of special data analysis techniques to enhance the signal-to-noise ratio (SNR) of the expected GW signals. Presently, the most credited algorithm is the well known matched-filtering technique, based on the correlations of the detector output with templates of the expected signal\cite{HELSTROM,PAPOULIS}. But, although very simple in principle, its practical application requires an exact theoretical knowledge of the shape of the expected signal, function of the unknown parameters describing the coalescing binary and the evaluation of the correlation of the detector output with several thousands of templates, requirements very difficult to satisfy for coalescing binary GW signals\cite{DHURANDHAR,SATHYAPRAKASH}.

The shape of the GW signal can be obtained by computing the gravitational radiation field generated by a system of two point-masses moving on a practically circular orbit. The solution of this problem requires the calculation of the gravitational radiation field to a very high order in terms of a post-Newtonian expansions, being coalescing compact binaries strongly dominated by relativistic effects. The large number of templates necessary for data analysis using matched-filtering technique arises problems due to the great computing power needed to perform this task on-line\cite{THORNE,DHURANDHAR}. In fact, as a consequence of the large band of interferometric detectors (some $kHz$), sampling rates of the order of $20\,kHz$ are foreseen, resulting in a huge amount of data/day to be analyzed on-line ($\approx 10$ GByte/day). Moreover, the computational cost depends on the number of parameters used for the phase approximation, on the accuracy of the sampling of the likelihood function (in connection with the capability of recovering weak signals) and on the chosen frequency band. For example, a direct application of the matched-filtering technique to the {\em VIRGO} antenna requires a computing power of at least $ \approx 0.3$ Teraflops starting from a minimum mass of $0.25 M_{\odot}$ with a SNR recovery of $90\%$, including the computing power necessary for the production of the templates\cite{VICERE,OWEN}. These estimates confirm that already at the stage of the first post-Newtonian correction large parallel computers are necessary for an on-line analysis of these signals. Of course, the analysis of such a large amount of information could be made off-line, but it is clear that great advantages may derive from a preliminary selection of the on-line data frames which may contain a GW signal, not last the fact that the selected data chunks will be object of more refined analysis, including also cross-correlations with data taken by other running interferometers.

Starting from the estimate of the huge computational cost of the data analysis for VIRGO, we decided to explore alternative approaches to try to solve this problem that can be used also in parallel to the existing ones. For this reason, also taking into account the relevance of a GW first detection, we decided to implement a hierarchical technique, based on fast on-line sub-optimal filters to be used as triggers (first level), in order to extract interesting data chunks, followed by an optimal technique like the matched filters one\cite{MILANO}. Our efforts were therefore aimed to the implementation of a rough analysis with the following characteristics: minimum signal losses with respect to the one using the matched Wiener-Kolmogorof filter, robustness against false alarm detection and, last but not least, need of low computational powers and possibility of running in real-time.

\section{IIR Adaptive Line Enhancer}
In a previous paper\cite{ACERNESE}, we tested an adaptive filter of the class of IIR (Infinite Impulse Response) Adaptive Line Enhancers (IIR ALE). The ALE algorithm is a technique designed to approximate the SNR gain obtained by the matched filter solution for a sinusoidal signal. The advantage of using ALE filters is that no a priori knowledge of the signal parameters (sinusoidal frequencies, amplitudes, phases or even the number of narrow-band components) is required. Therefore, center frequency band-pass filters can be used when the signal to detect is narrow-band and buried in a wide-band noise. And a slow frequency sweep very well approximate this case. In fact, instead of searching for the optimum Wiener Filter, that requires the knowledge of a general frequency response to be adapted to the signal, the basic idea of this procedure is that of starting from a band-pass response and then optimize its parameters. Something similar is done in Matched Filter technique, where the occurrence of a signal must be detected, in the hypothesis of known signal waveform, but with unknown parameters.

The optimization procedure may be carried out in the following way.
Let us suppose that the observed signal is $x_n=s_n+w_n$, where $s_n$ is a band-limited signal (eventually a single sinusoid) and $w_n$ is a white noise uncorrelated with $s_n$.
Let $Y(z)$ be the output of a digital filter with transfer function $H(z)$ in the Z-domain:
\begin{equation}
Y(z)=H(z)X(z)=H(z)[S(z)+W(z)]
\end{equation}
The error after filtering in the time domain will then be
\begin{equation}
d_n=y_n-s_n
\end{equation}
and in the Z-domain
\begin{equation}
D(z)=Y(z)-S(z)=H(z)W(z)+[H(z)-1]S(z).
\end{equation}

If signal and noise are assumed to be uncorrelated, referring to the above formula, the power spectrum of the error signal is
\begin{equation}
S_{d}(\omega )=\left| H(e^{j\omega })-1\right|^{2}S_{s}(\omega )+\left| H(e^{j\omega })\right| ^{2}S_{w}(\omega).
\end{equation}
The rms error can then be evaluated by inverting this formula, leading to
\begin{eqnarray}
E\left\{ d^{2}_n\right\} =\frac{1}{2\pi }\int\limits_{2\pi}\left| H(e^{j\omega })-1\right| ^{2}S_{s}(\omega )d\omega + \nonumber \\
\frac{1}{2\pi } \int\limits_{2\pi }\left| H(e^{j\omega })\right|^{2}S_{w}(\omega )d\omega.
\end{eqnarray}
Then choosing the root mean square error as the cost function to be minimized, the optimization will lead to the optimal filter transfer function $H(e^{j\omega })$.
Supposing that our simple model, consisting in a sinusoid immerse in gaussian noise, can be described in the frequency domain by
\begin{equation}
S_{s}(e^{j\omega })\cong 2\pi \delta (\omega -\omega _{0})  \quad \textrm{and} \quad  S_{w}(\omega )\cong \sigma _{w}^{2},
\end{equation}
then the root mean square error can be written as
\begin{equation}
E\left\{ d^{2}_n\right\} =\frac{1}{2\pi }\left| H(e^{j\omega_{0}})-1\right| ^{2}+\frac{\sigma _{w}^{2}}{2\pi}\int\limits_{2\pi }\left| H(e^{j\omega })\right| ^{2}d\omega
\end{equation}
which is minimized by the two conditions
\begin{equation}
\int\limits_{2\pi }\left| H(e^{j\omega })\right| ^{2}d\omega = \min ! \qquad \textrm{with }\, H(e^{j\omega _{0}})=1
\end{equation}
thus leading to a sharp band-pass filter.

On the basis of the above considerations, the idea embedded in a band-pass Adaptive Line Enhancer (ALE) is that if $H(e^{j\omega})$ is a parametric band-pass transfer function, described by one or more parameters,  then it is possible to find the best value of these parameters by calculating the minimum of the above error. If the analytical structure of the error function prevents from a closed form solution, we may rely on a numerical optimization both in the time and in the frequency domain. The above procedure is well suited to the ideal case, when all the parameters are well known. In the real case the signal to be detected has unknown or time varying parameters. Therefore, an adaptive implementation of the filter becomes necessary to detect the signal and to trace its time varying features.

\subsection{Optimization of ALE Parameters}
Suppose that the signal model is the sine wave plus (white) noise model
\begin{equation}
x_n=a \sin(\omega _{0}n+\theta )+w_n,
\end{equation}
input of an adaptive filter with a band-pass parametric structure. The optimization must be then performed looking for a maximum of its output power $\sigma _{y}^{2}$.
The performance function in this case is
\begin{equation}
E\left\{ \left| y_n\right| ^{2}\right\} =Z^{-1}(S_{y}(z))|_{n=0}
\end{equation}
Then since
\begin{equation}
Z^{-1}(S_{y}(z))=\frac{1}{2\pi j}\oint S_{y}(z)z^{n-1}dz
\end{equation}
then in the case of a linear filter with frequency domain response $H(e^{j\omega })$, the expected power of the filter output can be written as
\begin{equation}
E\left\{ |y_n|^2\right\} =\frac{1}{2\pi }\int\limits_{2\pi}S_{x}(\omega )\left| H(e^{j\omega })\right| ^{2}d\omega
\end{equation}
In the simple case of sine wave or band-pass signal
\begin{equation}
S_{s}(\omega)\cong \frac{a}{\sqrt{2}}\delta (\omega -\omega _{0})
\end{equation}
within uncorrelated white noise, $S_{w}(\omega)\cong \sigma _{w}^{2}$, we have
\begin{equation}
E\left\{ \left| y_n\right| ^{2}\right\} =\frac{a^{2}}{2}\left|H(e^{j\omega _{0}})\right| ^{2}+\frac{\sigma _{w}^{2}}{2 \pi} \int_{2 \pi} \left| H(j \omega) \right|^2 d \omega
\end{equation}
It is necessary to evaluate the performance function, in order to study the features of the performance surface of the system, and to define the most suitable  algorithm to reach the maximum. An unimodal surface will obviously simplify this search, avoiding the need for the use of a global search algorithm. Many implementations are reported in literature, characterized by different structures. Within this context two are the most relevant parameters: convergence speed and tracking capabilities, that are clearly contrasting requirements. An improvement may be achieved if the parameters of the optimization include both filter center band and the bandwidth, at least in the final stage of the convergence. This may improve the main feature of the method which is the capability to provide a running search of the center frequency, improving the detection of the instantaneous frequency.

\subsection{Choice of the prototype band-pass filter}
In principle any adaptive band-pass filter is suitable for our purpose. Anyway it may be convenient to properly shape the band in order to exactly match the shape of the signal in the frequency domain, since it will act as a weighting function in the frequency domain. Therefore, in this case it is convenient to use a filter prototype with the selected shape in the frequency domain, to which apply a parametric frequency transformation, in order to obtain the final band-pass filter.

We implemented the second-order band-pass Butterworth ALE filter\cite{RAJA} as
\begin{equation}
H(z)=(1-r^2)\left[ \frac{W_t z / (r+r^2) -1}{z^2-W_t z +r^2} \right]
\end{equation}
where $W_t=2r\cos(2 \pi f_0)$ is the only parameter dependent on the
center-frequency $f_0$, while $r$ is a fixed design parameter related 
to the filter bandwidth $B$ by the relation $B=F_s (1-r) /2$, where $F_s$ is the sampling frequency. 
The recursive algorithm for the filter can be expressed by:
\begin{eqnarray}
y_t &=& W_t \left( \frac{1-r}{r} \right) x_{t-1} - (1-r^2) x_{t-2} + 
W_t y_{t-1} -r^2 y_{t-2} \\
\alpha_t &=& \frac{\delta y_t}{\delta W_t} = \left( \frac{1-r}{r} 
\right) x_{t-1} + W_t \alpha_{t-1} +y_{t-1} -r^2 \alpha_{t-2} \\
R_{t+1} &=& \nu R_{t} + \alpha_t^2 \\
W_{t+1} &=& W_t + \frac{ \mu_a y_t \alpha_t }{R_{t+1}}
\end{eqnarray}
where $\nu \in (0,1)$ is the forgetting factor, introduced for the recursive computation of the normalizing factor $R$, $\alpha_t$ is the instantaneous gradient and $\mu_a$ is the scalar adaptation step size. 
Stability monitoring of these filters is very simple, being performed in order to verify the condition $|W_t|<2r$. 

Let us assume a coalescing binaries signal (chirp) injected in AWGN (Additive White Gaussian Noise) with a theoretical $SNR^2_{in}$ \footnote{The definition of signal to noise ratio we used is $SNR^2_{in}=\frac{2E}{\sigma^2 F_s}$ where $E=\sum_i x^2_i(t)$ is the signal energy, $\sigma^2$ is the noise variance and $F_s$ is the sampling rate.}. The output of {\em ALE} filter has an improvement of
\begin{equation}
\label{eq:imp}
I=\left(\frac{SNR^2_{ALE}}{SNR^2_{in}} \right) = 
\frac{\int |H(z)|^2 \Phi_{ss} (dz/z) }{2 \pi j \sigma^2_n B_{f}}
\end{equation}
where $\Phi_{ss}$ is the power spectral density of the signal, $\sigma_n^2$ is
the input noise power and $B_f$ is the equivalent noise bandwidth.
When the ALE filter tracks the input signal, $f_0$ tends to coincide with
the signal instantaneous frequency and, therefore, the equation~\ref{eq:imp} can be written as\cite{MILANO}
\begin{equation}
\label{eq:improve}
SNR^2_{ALE}=  SNR^2_{in} \cdot \frac{(1+r)}{2(1-r)}
\end{equation}
From equation~\ref{eq:improve} we can state that, if the condition $(1-r)<<1$ is satisfied and ALE filter closely tracks the signal, there is an improvement of $SNR^2_{in}$ by a factor of $(1+r)/(2(1-r))$ simply because ALE is a band-pass filter with a adaptive center frequency. 

\subsection{ALE thresholds: probability of false alarm and probability of detection}
Using an ALE filter for detection, a very delicate problem is the choice of the threshold $T$, to discriminate between the alternative hypotheses of absence or presence of a signal at an assigned level of confidence.
Since we use an ALE filter, we can assume that we have an approximation of the instantaneous signal amplitude, while the signal phase is unknown. 
In these hypotheses, in presence of only \emph{gaussian} noise with variance $\sigma^2_n$, the output power of the ALE filter is $\sigma^2_N=2 \sigma^2_n (1-r)/(1+r)$, and it can be modeled with a random variable having a \emph{Rayleigh} distribution with spread parameter $\sigma^2_N$\cite{HELSTROM}: 
\begin{equation}
\label{eq:Rayleigh}
p_0(x)= \frac{x}{\sigma_N^2} \exp \left\{-\frac{x^2}{2 \sigma_N^2} \right\}
\end{equation}
If a signal is present and is tracked, then the output power statistic becomes a \emph{Rice} distribution centered on the signal energy $\cal E$\cite{HELSTROM}, and with spread parameter $\sigma_S ^2=2\sigma^2_n(1-r)/(1+r) = \sigma^2_N$:
\begin{equation}
\label{eq:Rice}
p_1(x)= \frac{x}{\sigma _S^{2}}I_{0}\left(\frac{x\cal E }{\sigma _S^{2}}\right) e^{-\frac{x^{2} +{\cal E}^2}{2{\sigma_S^2}}} \qquad \textrm{with } x>0
\end{equation}
where $I_{0}(\cdot)$ is the modified $0^{th}$ order Bessel function of first kind. 
The false alarm probability, assuming independent samples, can be written as
\begin{equation}
P_{fa} = \int_T^{+\infty} p_0(x) dx = \exp \left\{-\frac{T^2}{2 \sigma_N^2} \right\}
\end{equation}
that is a suitable closed form for the evaluation of the optimum detection threshold $T$ that can be calculated as
\begin{equation}
\label{eq:soglia}
T=\sqrt{-2 \sigma_N^2 \log P_{fa}}= \sigma_n \sqrt{-4 \left( \frac{1-r}{1+r} \right) \log P_{fa}}
\end{equation}

Concerning the probability of detection $P_d$, assuming that the signal is present and tracked, then we have
\begin{equation}
\label{tt1}
P_d=\int_T^{+\infty} p_1(x) dx ={\cal Q}\left( SNR_{ALE},T/ SNR_{ALE} \right)
\end{equation}
where $\cal Q$ is usually referred to as the Marcum ${\cal Q}$ function
\begin{equation}
{\cal Q}(a,b) = \int_b^\infty x \, e^{-\frac{a^2+x^2}2} \, I_0(ax) \, dx
\end{equation}

\begin{figure}[htb]
	\begin{center}
	\begin{tabular}{c}
				\includegraphics[width=10cm]{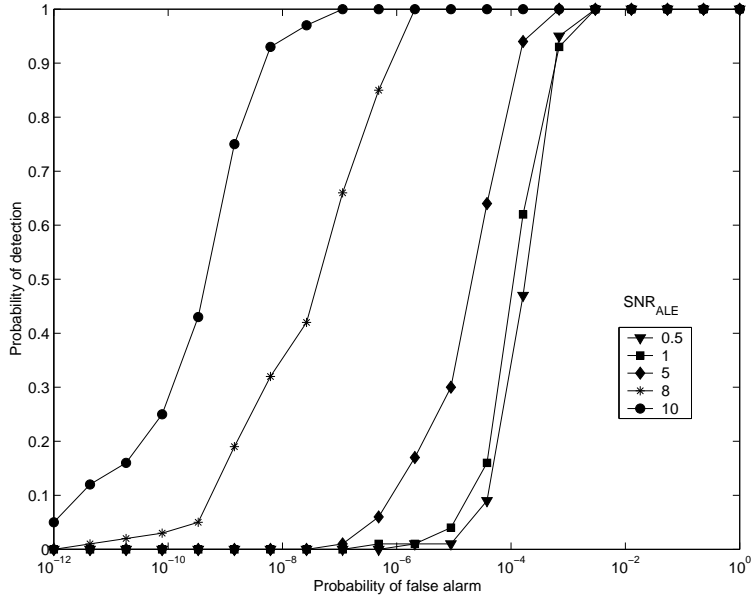}
			\end{tabular}
		\end{center}
				\caption[SingleAle]{\label{fig:ALE}The empirical ROC curve of single ALE filter, evaluated at different $SNR_{ALE}$, is shown.}						
\end{figure}

\begin{figure}[htb]
	\begin{center}
	\begin{tabular}{c}
					\includegraphics[angle=-90,width=12cm]{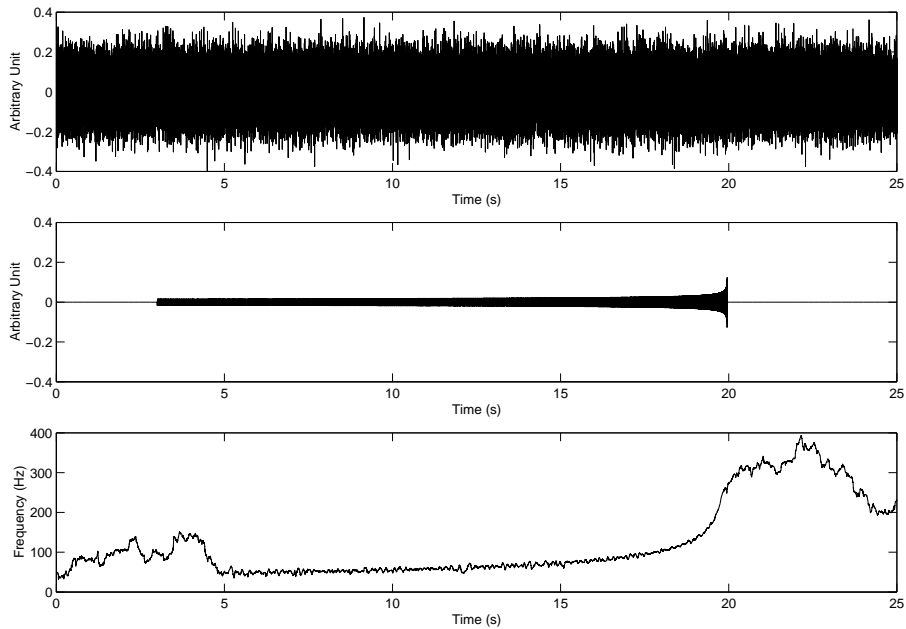}
			\end{tabular}
	\end{center}
				\caption[AleTest]{\label{fig:ALETest}An example of Ale filter output is shown. From the top, a data set of binary signal injected in gaussian noise with $SNR_{ALE}\cong10$, the binary signal and the frequency estimated by ALE filter, are shown.}
\end{figure}

We made several synthetic experiments to test the proposed algorithm. We used the waveform generated by coalescing binary stars with masses $1.4 - 1.4 M_{\odot}$ at the $2.5$ post newtonian expansion. The noise source has a Gaussian probability distribution in the frequency band $50\div5000\,Hz$. The ALE filter has a starting frequency of $50\,Hz$ and a fixed band of $25\,Hz$ corresponding to $r=0.990$. The empirical ROC curve of the ALE filter, necessary to characterize the performance as a sub-optimal detector, is shown in figure~\ref{fig:ALE}.
It is immediately clear that a high probability of detection within a high false alarm probability is obtained also at very low signal-to-noise ratio, demonstrating, in this way, the feasibility of using an IIR ALE filter as trigger with an assigned confidence level. In fact, choosing a higher false alarm probability, we may use an ALE IIR like an efficient trigger also at very low SNR.
In figure~\ref{fig:ALETest} an example of ALE filter output is shown: from the top it can be seen a data set of binary signal injected in gaussian noise with $SNR_{ALE}\cong10$, the binary signal, and the frequency estimated by ALE filter.

\subsection{Parallel implementations}
An accurate analysis of the ALE performances shows that in the low SNR case only the choice of a starting center frequency of the ALE filter very close to the frequency to detect may allow taking advantage of the gradient procedure thus reducing the effect of high level noise. In our case, although the transition of the gravitational wave chirp through all the frequencies of the sensitivity band of the interferometric detectors is theoretically predictable, the time at which the gravitational wave chirp passes through the starting center frequency of the ALE filter is not predictable.
Actually this may be a problem for the ALE filter. In fact, when a GW chirp reaches the design starting center frequency of the ALE filter, if its present center frequency has moved from its design one due to the input noise effects, then the probability of detection largely decreases.

In order to keep this probability the highest possible, closest to the maximum theoretical one, it is important to {\em synchronize} the design starting center frequency of the ALE filter with the GW signal frequency. A possible solution to this problem is to use parallel techniques, like the one of building a bank of ALE (band-pass) filters, with a common single design starting center frequency, but starting at different times, sequentially and with fixed delay times. The number of ALE filters of the bank is defined at the beginning of the procedure. When the bank is filled, and a new ALE filter is started, then the oldest one is eliminated.

The basic idea on which this procedure is based on, is clearly that of overcoming the lack of synchronization among the ALE filters at start-up and the possible presence of a GW signal. Of course, the shortest is the delay time, the best is the synchronization. In the limit it is possible to make this delay time coincident and synchronous with the inverse of the sampling frequency of the data acquisition system. In fact, due to the low computational power required it is possible also to think of a bank of millions of ALE filters, running in parallel for a real-time data analysis.
This solution allows a faster identification of the frequency to be detected, coupled with a more robust signal identification and tracing. In fact, if the ALE filters are very close each other, then in presence of a GW signal it is obvious that a quite large number of them should detect the signal, allowing also to perform a better detection. What need to be still to be completed is the development of a real-time decision unit, necessary to statistically interpret the data from the parallel filter bank and to control the center filter frequencies.

\begin{figure}[htb]
	\begin{center}
		\begin{tabular}{c}
					\includegraphics[angle=-90,width=12cm]{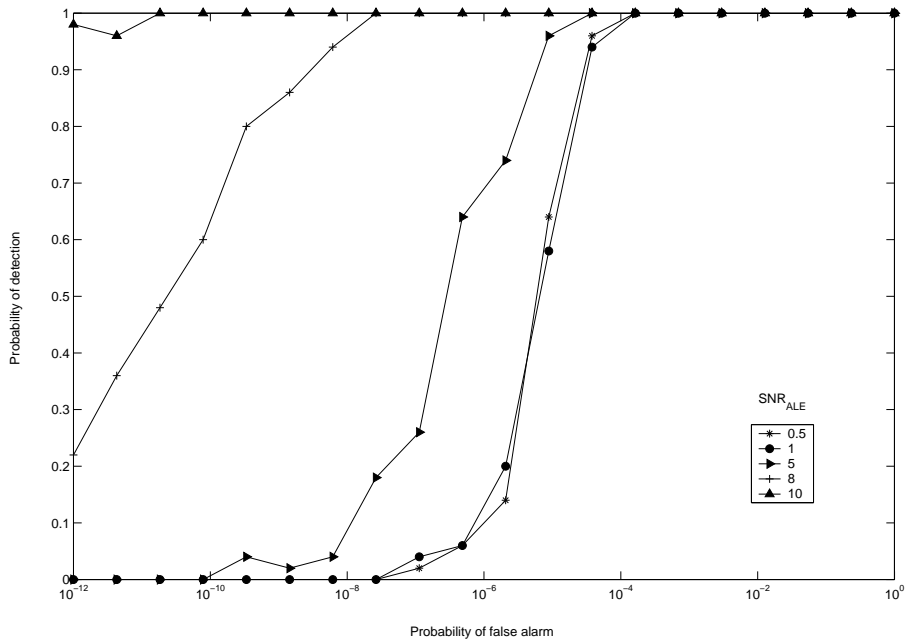}
			\end{tabular}
				\end{center}
				\caption[MultiAle]{\label{fig:MultiAle}The empirical ROC curve of Parallel ALE filter, evaluated at different $SNR_{ALE}$, is shown.}
\end{figure}

All the numerical experiments carried on by us, in low SNR conditions, exploited this parallel approach, and produced very good results. The empirical ROC curve of the parallel ALE filter, necessary to characterize the performance as a bank of sub-optimal detectors, is shown in figure~\ref{fig:MultiAle}. As it can be seen, an improvement of the performances is archived, respect to the single ALE filter. 

Finally, for completeness, a thoroughly parallel structure could also be used in order to trace the error surface, for example by sweeping over the entire frequency range, thus avoiding possible looping on local minima, and then providing robustness to the detection.

\section{Conclusions and future improvements}
We have analyzed and discussed the performances of an IIR ALE filter to perform real-time triggering on the output data from interferometric antenna, like the VIRGO one, showing that it is suitable for GW data analysis from coalescing binaries. This filter is based on adaptive modification of a band-pass filter parameters, changed according to a suitable cost function, within the hypothesis that the signal to detect is narrow-band (even if the center band is changing along the time) buried in a wide-band noise. For this task we performed some choices, analyzing the performances of $2^{nd}$ order direct implementation of IIR adaptive filters, either in the case of a single filter or in the case of a parallel bank of filters. From the analysis of the detection probability, we obtained the results that are shown in figure~\ref{fig:ALE} and~\ref{fig:MultiAle}.

The simulations we performed demonstrate that on-line triggering on gravitational wave interferometric experiments is feasible without a huge or complex computing power and without the necessary a priori knowledge of the signal waveform. Coupling a parallel process of different algorithms will surely provide a better and robust on-line tagging of binary coalescence signals, although further studies are necessary to test in greater detail the parallel version of ALE triggers.


\end{document}